\documentclass[10pt]{iopart}

\usepackage{iopams}  
\usepackage{hyperref}
\usepackage{graphicx}
\usepackage{color}

\begin{document}

\title[Material Ratio Parameter Uncertainty]{Procedure to Approximately Estimate the Uncertainty of Material Ratio Parameters
due to Inhomogeneity of Surface Roughness}

\author{Dorothee H{\"u}ser$^1$, Jonathan H{\"u}ser$^3$, Sebastian Rief$^2$, J{\"o}rg Seewig$^2$, Peter Thomsen-Schmidt$^1$}

\address{1 Physikalisch-Technische Bundesanstalt, Bundesallee 100, 38116 Braunschweig, Germany}
\address{2 Institute for Measurement and Sensor Technology, University of Kaiserslautern, Gottlieb-Daimler-Stra{\ss}e, 67663 Kaiserslautern, Germany}
\address{3 Software and Tools for Computational Engineering, RWTH Aachen University, Seffenter Weg 23, 52074 Aachen, Germany}
\ead{dorothee.hueser@ptb.de}

\begin{abstract}
Roughness parameters that characterize contacting surfaces with regard to friction and wear are
commonly stated without uncertainties, or with an uncertainty only taking into account a very 
limited amount of aspects such as repeatability of reproducibility (homogeneity) of the specimen. This makes it
difficult to discriminate between different values of single roughness parameters.

Therefore uncertainty assessment methods are required that take all relevant aspects into account. In
the literature this is scarcely performed and examples specific for parameters used in friction and wear
are not yet given.

We propose a procedure to derive the uncertainty from a single profile employing a statistical method that is
based on the statistical moments of the amplitude distribution and the autocorrelation length of the profile.
To show the possibilities and the limitations of this method we compare the uncertainty derived from a single profile
with that derived from a high statistics experiment.
\end{abstract}

\pacs{06.20.Dk,06.30.Bp}
% 06.20.Dk Measurement and error theory
% 06.30.Bp Spatial dimensions (e.g., position, lengths, volume, angles, and displacements) - Metrology, measurements, and laboratory procedures
\vspace{2pc}
\noindent{\it Keywords\/}: surface roughness, bearing/material ratio, stochastic process, uncertainty
\submitto{\MST}
\maketitle
% um die Abbildungen anzupassen, fuer Refereeprozess dann single col und 12pt
%\ioptwocol
%

\section{Introduction}
Surface roughness is a relevant feature for contacting surfaces besides material properties such as stiffness and
adhesion. Whether regarding joints and bearings in mechanical engineering, medical prostheses, or the contact between a cutting
edge and the work piece, the surfaces in contact need characterization and assessment accordingly.

Roughness is a stochastic property that is characterized by a variety of statistical estimators delivering measures to
parameterize the height distribution, the distributions of slopes, of vertical and lateral peak-valley sizes. Furthermore
autocorrelation length, fractal dimension, and many more quantities are used to quantify stochastic features of a topography,
hence roughness. To quantify roughness of contacting surfaces in particular a set of parameters derived from the so-called
bearing ratio distribution is defined in ISO 13565-2 and presented in detail in
Bushan \cite{BBushan2001} % section 2.2.2.4 
and in Whitehouse \cite{Whitehouse1994ch2}. % section 2.1.1.1. 

Rough topographies are asperities and dales of randomly distributed sizes and shapes. Their mountainous structure shows
an autocorrelation with average autocorrelation lengths. Comparable to a regular sampling on periodic structures,
both the bandwidth and the resolution of the sampling process play a role for textures that have similar,
repetitive features.
The spatial resolution of the measurement process is the measure of how closely structures can be resolved, which includes
the size of the area of a surface over which the mapping or probing instrument integrates. This means that a sampled height value 
is not the height of a \textsl{point} but the average or maximum height of an \textsl{area}. The bandwidth of a sampling process 
is characterized by the distance of neighboring sampling points and the width and shape of the impulse function of the
sampling train limiting the high frequency resolution and possibly causing aliasing effects.
Additionaly, it is characterized by the total sampling area limiting the maximum wavelength and possibly biasing
the autocorrelation characteristics.

If roughness measurement instruments do not supply an uncertainty estimate of the roughness parameters,
they do not state a complete measurement result. Uncertainties can be stated that are caused by the measurement
process of the instrument, if the instrument is well understood by the user or manufacturer.
A manufacturer of an instrument, however, cannot implement {\`a} priori knowledge on the characteristics of
the measurement objects of his customer. The problem to solve is to join contributions of the instrumental's intrinsic
stochastic processes and the measurement object's characteristics, the inhomogeneity of
its micro topography, to the uncertainty.

A procedure for estimating an uncertainty of roughness parameters was proposed by Haitjema
for tactile profilers \cite{Haitjema2013} and in a more general sense \cite{Haitjema2015}.
It is common to claim traceability of a roughness instrument when it is calibrated using test
objects with deterministic topographies. Some of them are
uniform grids of defined shape, such as triangular or sinusoidal, others are apparently random profiles, but are
manufactured as deterministic predefined function that is repeated in a systematic way.

The uncertainty of roughness parameters of deterministic topographies depends on the measurement principle in the sense of
the above mentioned sampling bandwidth \cite{LeachHaitjema2010}, the uncertainty induced by the instrument itself (noise, quantization,
% /home/hueser01/iPadTransfer/roughness_characterize/lit_rough/LeachHaitjema-Bandwidth_characteristics_and_comparisons_of_surface_texture_measuring_instruments-2010.pdf
stability, positioning / geometric deviations, cross talk, calibration etc.) and on the
evaluation method, i.e.\ the filtration, the algorithm to determine the parameter and the numerical realization of the
implementation of both of them \cite{Haitjema2013}. To evaluate the uncertainty contributions of the measurement
instruments' components, the stochastics of their error influences are carried out partially as an uncertainty budget,
while special aspects such as noise are simulated by Monte Carlo methods as virtual instrument
\cite{HaitjemaMorel2000,XuDziombaKoenders2011,Giusca2011}.
% Dokumente/pubs_literat_all/Literatur/virtuelle_MessGeraete/XuDziombaKoenders-Modelling_and_simulating_scanning_force_microscopes_for_estimating_measurement_uncertainty-a_virtual_scanning_force_microscope-2011.pdf
% qspm.bib
% /home/hueser01/Dokumente/1-Algos_devel/bosse_lit_rough_stepheight/GiuscaLeachForbes-A_virtual_machine-based_uncertainty_evaluation_for_a_traceable_areal_surface_texture_measuring_instrument-2011.pdf
% surftec.bib

\begin{figure}
\begin{center}
\includegraphics[width=83mm,keepaspectratio]{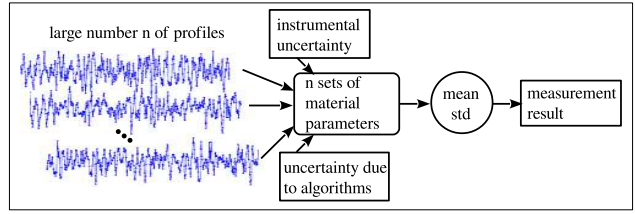}
\end{center}
\caption{According to the vocabulary of metrology the measurement result requires to state an uncertainty in addition to
  the physical quantity itself. The uncertainty usually is obtained from sufficient statistics, taken from
  a large number of profiles on one surface.}
\label{fig:RkparamsPrinciple1}
\end{figure}
For non-deterministic, i.e.\ stochastic topographies, a major contribution to the uncertainty of roughness parameters
besides instrumental limits is its inhomogeneity, the variation of the topography itself.
It is the interrelation between bandwidth limit of
probing, relocation of samples with size, correlation lengths, periodicities, and randomness of the structures and features
of the topography. Therefore performing a Monte Carlo simulation of the instrumental contributions 
without considering the interactions with the object only gives
a component of the error budget that may be significantly smaller than the topography induced contribution.
There is a strong demand for modeling the surface topography as well.
A sufficiently representative set of profile resp.\ areal scans of the appropriate bandwidth are required to
assess the texture characteristics of a surface. An uncertainty assessment must be made in addition to stating a
measured quantity as \textsl{measurement result}
in the sence of the international vocabulary of metrology, which states that a measurement result is generally expressed
as a  measured quantity value and a measurement uncertainty. For the aspect of surface inhomogeneity, a larger number of values
for each of the roughness parameters is required. Let $R_{\mathrm{x}, i}$ be one of the roughness parameters
%\footnote{In contrast to profilometric roughness ISO standards, we prefer the subscript notation as employed in
%areal roughness tandard ISO 25178-2 to make mathematical equations better readable.}
of profile $i$, then the mean is estimated by $\bar R_\mathrm{x} = \frac{1}{n} \, \sum_{i=1}^n R_{\mathrm{x}, i}$
and the standard deviation
$s\colon (R_{\mathrm{x}, 1},\dots, R_{\mathrm{x}, n}) \, \mapsto \, s(R_{\mathrm{x}, 1},\dots, R_{\mathrm{x}, n})$ 
by $s(R_\mathrm{x}) = \sqrt{\frac{1}{n-1} \, \sum_{i=1}^n (R_{\mathrm{x}, i} \, - \, \bar R_\mathrm{x})^2}$.
Then mean and standard deviation of the roughness parameters can be evaluated to
obtain a \textsl{measurement result} as illustrated in Fig.~\ref{fig:RkparamsPrinciple1}.
Regarding nowadays computer technology and comparing it to instrumentation,
it is often the case that simulations are faster and less expensive than measurements. In case of tactile instruments the
measurement process may cause wear or even damage. Therefore, Monte Carlo simulations may be preferred, if there
exists {\`a} priori knowledge of the statistical behavior of the data for deriving simulation results from the data with
insufficient empirical statistics. In addition to the statistical analysis of the topography influence, the uncertainties caused
by the instrumental devices as well as those caused by the choice of the algorithmic procedures, i.e.\ the filtration methods
\cite{HueserThSMees2016} and the way of evaluating the Abbott curve, contribute to the final result.
\begin{figure}
\begin{center}
\includegraphics[width=83mm,keepaspectratio]{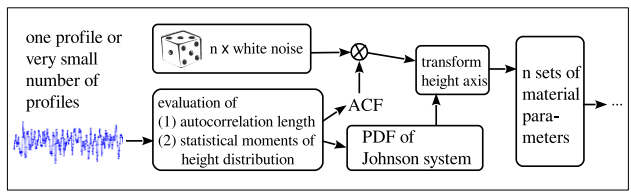}
\end{center}
\caption{With {\`a} priori knowledge of the statistical behavior of the data, the uncertainty could approximately
 be estimated by deriving simulation results from the data with insufficient statistics.}
\label{fig:RkparamsPrinciple2}
\end{figure}

For more than fourty years a variety of models to describe and simulate the roughness of surface topographies have already
been developed employing approaches of random field theory, time series, and non-causal stochastic processes \cite{WuAli1978}.
Wu \cite{JJWu2000} compares the approaches to convolve white noise with appropriate weight functions
% /home/hueser01/Dokumente/pubs_literat_all/pubs_reports/25_RkUncertainty/lit/Wu-Simulation_of_rough_surfaces_with_FFT-2000.pdf
that are either obtained by autocorrelation resp.\ the power spectrum density \cite{OgilvyFoster1989}
% /home/hueser01/Dokumente/1-Algos_devel/Synthetic_Rough_for_MC/OgilvyFoster-Rough_surfaces-gaussian_or_exponential_statistics-1989.pdf
or by auto regressive (AR) models \cite{Seewig2000ch53},
% /home/hueser01/Dokumente/1-Algos_devel/Synthetic_Rough_for_MC/ARmodel/Uchidate-Generation of reference data of 3D surface texture_using_the_non_causal-2D_AR_model-2004.pdf
% /Dokumente/pubs_literat_all/Literatur/Theorytools/Dissertation_Seewig.pdf
or the approach to use the power spectrum and instead of multiplying it with the Fourier transform of white noise,
to multiply it with uniformly random distributed phases. Uchidate has investigated non-causal AR-models 
for surface topographies \cite{Uchidate2004}, because space is not restricted to causality as time does.
A method of obtaining random data while maintaining the correlation properties was
given by Theiler \textsl{et al.} \cite{Theiler1992}, which has applied to roughness measurements by Morel
\cite{MorelDiss2006}.

Many of the roughness models presume that surface heights are normally distributed, i.e.\ that
they have a Gaussian shaped amplitude distribution. The deviation from this presumption is quantified by
the roughness parameters $R_\mathrm{sk}$, which is the \textsl{skewness}, i.e. the third
statistical moment, and $R_\mathrm{ku}$, which is the \textsl{kurtosis} or excess, i.e.\
the fourth statistical moment, of the probability density function of topography height values.
To parameterize non-Gaussian probability density distributions such that the distributed quantity is
transformed to a quantity that then has a Gaussian probability density distribution, a system of
functions has been introduced by Johnson in 1949 \cite{HillHillHolder1976}. To estimate the appropriate function of that set with
its parameters accordingly, Hill, Hill, and Holder \cite{HillHillHolder1976} have developed an algorithm in 1976 that we are
employing for our suggested procedure. For more than fifteen years, the Johnson system has been applied to roughness analysis
and simulation \cite{ChilamakuriBhushan1998,Bakolas2003}.
% /home/hueser01/Dokumente/1-Algos_devel/Synthetic_Rough_for_MC/Bakolas-Numerical_generation_of_arbitrarily_oriented_non-Gaussian_3D_rough_surfaces-2003.pdf

Stochastic data require huge samples for statistical analysis and assessment. In quality assurance in industrial life, however,
small representative samples are drawn to spot-check a process or pieces. Therefore, we have investigated, how well the
uncertainty of material ratio parameters of roughness data can be estimated from a single measurement, one profile or a
single area scan. In the next section, the definition according to ISO standards of material ratio parameters will be presented
and the ambiguities of the definition will be discussed. Section three deals with the influence of sampling effects on the
autocorrelation function \textsl{ACF} and on the probability density function \textsl{PDF} of a topography revealing the sampling effect
by looking at synthetic, well defined topographies, defined by Fourier series. In section four, we will give details on the
probability density distributions that are useful to describe amplitude distributions of roughness profiles.
Section five is dedicated to show a way for an approximate estimation of the inhomogeneity component of the 
uncertainty of the material ratio parameters by deriving Monte Carlo simulated profiles from a small set of
profiles or even a single profile as depicted in Fig.~\ref{fig:RkparamsPrinciple2}.
The procedure is a coarse guess being helpful for industrial processes,
but does not preempt from taking large data samples to obtain reliable statistical results for research purposes.

The proposed procedure is based on investigations on simulated synthetic profiles of known Fourier series components
as well as experimental profiles of a tactile areal profiler, a custom built micro topography
measurement system \cite{ThomsenS2011}.
The vertical axis of the measurement system is realized by a stylus with its vertical
movement measured interferometrically directly in line with its probe tip, i.e.\ without Abb{\'e} offset
and without any arc error.
The stylus is guided by an air pressure bearing and its probing force is controlled by magnetic fields.

The experimental data have been taken on different kind of industrial surfaces.
The experiments were carried out
on an area of $4 \times 4 \; \mathrm{mm}^2$ on surfaces with $R_\mathrm{a}$ values lying in an interval of
$(0.1, \, 2] \; \mathrm{\mu m}$ according to ISO 4288:1996/Cor 1:1998.
The filtration\footnote{The filtration methods are defined in ISO standards ISO 16610-21:2011, ISO 16610-22:2015,
ISO/TS 16610-28:2010, and in ISO/TS 16610-31:2010 for profiles and furthermore for areal scans in
ISO 16610-61:2015. ISO 16610 parts 21 and 28 replace ISO 4287 and ISO 16610 parts 31 (and 28) and replace ISO 13565-1.}
cuts off the waviness contribution by wavelength $\lambda_\mathrm{c}$ and
the high frequency contribution by wavelength $\lambda_\mathrm{s}$ to suppress noise,
to reduce apparent low frequencies induced by the folding of frequencies around the sampling
frequency \cite{Whitehouse1994ch3},
% section 3.1.1, 
and to match instrumental bandwidths \cite{LeachHaitjema2010}.
 ISO 4288 defines the choice of the cut off
wavelengths $\lambda_s$ and $\lambda_c$ according to the amplitude parameters $R_\mathrm{a}$ and $R_\mathrm{z}$.
ISO 3274 defines the maximum allowed width of sampling intervals according to the cut off wavelengths, for
our experimental setting this is $\Delta x \, \le \, 0.5 \, \mathrm{\mu m}$ and the radius of the probe tip
is $R \, \approx \, 2 \, \mathrm{\mu m}$.
That means that current ISO standards define the choice of the bandwidth according to amplitude parameters rather
than correlation length and other horizontal parameters, an issue that will also be discussed in section three.

\section{Definition of Material Ratio Parameters}
% /home/hueser01/Dokumente/1-Algos_devel/Abbott_parameters/Abbott_doku/defMaterialAnteil2.svg

\begin{figure}
\begin{center}
\includegraphics[width=80mm,keepaspectratio]{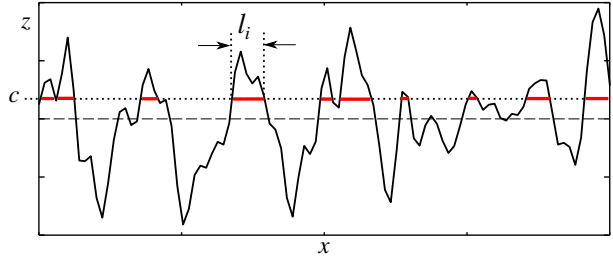}
\end{center}
\caption{Definition of the material ratio.}
\label{fig:DefinitionMr}
\end{figure}

\begin{figure}
\begin{center}
\includegraphics[width=80mm,keepaspectratio]{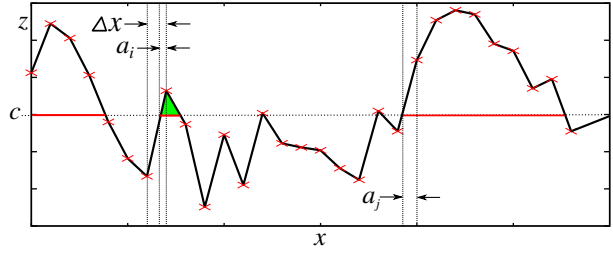}
\end{center}
\caption{Material ratio with intersections between profile asperities and horizontal line at height level $c$.}
\label{fig:DxWichtung}
\end{figure}

To clearify the relation between the statistical height distribution of a surface and the roughness parameters
that are used to characterize surface contact, this section presents the definition of the material ratio
parameters in detail.
Abbott and Firestone have proposed to describe the area of contact between surfaces by characterizing the area
of each surface as the ratio of air to material at any level $c$. The parameter \textsl{material ratio} $Mr$, also
called \textsl{bearing ratio}, is a function of height level $c$ \cite{BBushan2001}.
Let $L$ be the length of the total profile, then the sum of the
length pieces $l_i$ intersecting the asperities at level $c$, i.e.\ $l_i(c)$, delivers $Mr  \equiv r_\mathrm{M}$
\begin{equation}
r_\mathrm{M} (c) \; = \; \frac{1}{L} \, \sum_i l_i(c)
\label{eqDefinitionMr}
\end{equation}
as illustrated in Fig.~\ref{fig:DefinitionMr}. As a double letter identifier is inappropriate for
maths formulae, we denote the material $r_\mathrm{M}$ rather than $Mr$. The inverse of the
function material ratio depending on height level, i.e.\ the distribution $c$ vs.\ $r_\mathrm{M}$,
is called \textsl{Abbott-Firestone distribution}, abbreviated \textsl{Abbott-curve}.
The algorithm to evaluate all $l_i(c)$ from intersecting asperities and subsequently calculating each intersection between
the horizontal line at $z = c$ and the asperities has a complexity that can be avoided if the residual error
\begin{equation}
\Delta r_\mathrm{M} (c) \; = \; \sum_i \left( \frac{1}{2} \Delta x \, - \, a_i \right)
\end{equation}
is sufficiently small.
The statistical distribution of negative differences $0.5 \Delta x \, - \, a_i$ is similar
to that of positive, so that they almost cancel on average in most cases.
The distances $a_i$ are those between the intersection of an asperity surface and the neighboring knot of the profile
as illustrated in Fig.~\ref{fig:DxWichtung}.

A fast and efficient approach is to sort all discrete height values
of the equidistantly sampled profile according to their values $z_i$
\begin{equation}
z_{i_1} \, \ge \, z_{i_2}  \, \ge \dots \, \ge \, z_{i_k}  \, \ge \dots \, \ge \, z_{i_n} 
\label{zSort}
\end{equation}
such that with $L \, = \, (n-1) \, \Delta x$
\begin{equation}
r_\mathrm{M} (c = z_{i_k}) \; \approx \; \frac{\Delta x}{L} \, \sum_{\nu = 1}^{k} 
\left(1 \, - \, \frac{1}{2}\delta_{i_\nu,1} \, - \frac{1}{2}\delta_{i_\nu,n}\right)
\end{equation}
with $(x_1, z_1)$ and $(x_n, z_n)$ being the border positions of the original profile and
$\delta_{i_\nu,1}$ and $\delta_{i_\nu,n}$ denoting the Kronecker symbols to treat the border positions
appropriately.

Furthermore we approximate this by
\begin{equation}
r_\mathrm{M} (c = z_{i_k}) \; \approx \; \frac{k-0.5}{n} .
\label{eqSortMr}
\end{equation}
Avoiding the values $r_\mathrm{M} = 0$ and $r_\mathrm{M} = 1$ is required if the inverse error function
$\mathrm{erf^{-1}}$ to parameterize the relation $c$ vs.\ $r_\mathrm{M}$ of profiles with Gaussian distributed
height values is used, which is fulfilled by using $k - 0.5$ rather than $k$.
% Our experience applying both methods to a variety
% of different types of profiles has verified this suggestion. PTB roughness reference software RPTB V2.0ff
% provides both algorithms, that of Eq.~(\ref{eqDefinitionMr}) and that of Eq.~(\ref{eqSortMr}).

The international standard ISO 13565-2 defines a set of $5$ parameters derived from the \textsl{Abbott-curve} for
profiles and ISO 25178-2 the corresponding parameters for areal scans:
\begin{itemize}
\item the core height $R_\mathrm{k}$, which is the distance between the highest and lowest level of the core profile
resp.\ for areal data $S_\mathrm{k}$ of the core surface,
\item the reduced peak height $R_\mathrm{pk}$ and reduced valley/dale height $R_\mathrm{vk}$, which
are the height of the protruding peaks above the core profile after reduction process in case of $R_\mathrm{pk}$ and the
height of the protruding dales below the core profile after reduction process in case of $R_\mathrm{vk}$.
For areal scans they are the height of protruding peaks above resp.\ dales below the core surface
and again the identifier $R$ is replaced by $S$,
\item the two material ratio quantities giving once the ratio of the area of the material at the intersection
line which separates the protruding hills from the core profile resp.\ surface to the evaluation area, shortly
named peak ratio $\mathrm{Mr}1$ resp.\ $S_{mr1}$; secondly the ratio of the area of the material at the intersection
line which separates the protruding dales from the core profile resp.\ surface to the evaluation area, shortly
named peak ratio $\mathrm{Mr}2$ resp.\ $S_{mr2}$.
\end{itemize}
The core height $R_\mathrm{k}$ is the negative slope of a regression line within a $40 \, \%$ interval for the
core material. The $40 \, \%$ interval $[r_\mathrm{M}(c_\mathrm{p}), r_\mathrm{M}(c_\mathrm{v})]$ with 
$$
r_\mathrm{M}(c_\mathrm{v}) := r_\mathrm{M}(c_\mathrm{p}) + 0.4
$$
is chosen such that the slope of the secant takes a minimum:
\begin{equation}
  \min_{c_\mathrm{p}, c_\mathrm{v}} \left\{ \frac{c_\mathrm{p} \, - \, c_\mathrm{v}}{r_\mathrm{M}(c_\mathrm{v}) \, - \, r_\mathrm{M}(c_\mathrm{p})}
  \right\}
\label{minSecant}
\end{equation}
In case of smooth \textsl{Abbott-curves} this interval coincides well with the interval of minimum slope $R_\mathrm{k}$,
but not in any case. The search of the minimum secant rather than slope has been chosen at times when computer time
was more costly and processing was slow.
Furthermore, a discrete set of height values with $c_\mathrm{v}$ at a position for $k = \mathrm{v}$, i.e.\ with
$z_{i_\mathrm{v}}$ and $c_\mathrm{p} = z_{i_\mathrm{p}}$ rather than a continous $40 \, \%$ interval is used.
Consequently the interval is a discretization
$$
 r_\mathrm{M}(z_{i_\mathrm{v}}) \, - \, r_\mathrm{M}(z_{i_\mathrm{p}}) \approx 0.4
$$
approximating $[r_\mathrm{M}(c_\mathrm{p}), r_\mathrm{M}(c_\mathrm{v})]$.

If the regression line fitted to the \textsl{Abbott-curves} within
$[r_\mathrm{M}(c_\mathrm{p}), r_\mathrm{M}(c_\mathrm{v})]$ is given by
\begin{equation}
  z (r_\mathrm{M}) \; = \; - R_\mathrm{k} \, r_\mathrm{M} \; + \; c_1 ,
\end{equation}
the parameter $R_\mathrm{k}$ is called core height.

For an amplitude distribution of sample size $n \rightarrow \infty$ and if it is a Gaussian, the \textsl{Abbott curve}
is the inverse error function $\mathrm{erf}^{-1}$, thus the negative slope at the position $r_\mathrm{M} = 0.5$ is
$\sqrt{2 \pi} \, R_\mathrm{q}$. The negative of the slope of a regression line to
$\mathrm{erf}^{-1}$ for $r_\mathrm{M} \in [0.3, 0.7]$ is $R_\mathrm{k} \cong 2.5739 \, R_\mathrm{q}$
\cite{Seewig2013}, which is greater than the slope at the $50 \%$-position with $\sqrt{2 \pi} \cong 2.5066$.

If $c_2 \, = \, c_1 \, - \, R_\mathrm{k}$ the ratio parameters $\mathrm{Mr}1 \equiv r_1$ and $\mathrm{Mr}2 \equiv r_2$
are obtained via the inverse \textsl{Abbott-curve}
\begin{equation}
r_1 \; = \; r_\mathrm{M}(c_1) \qquad \mathrm{and} \qquad r_2 \; = \; r_\mathrm{M}(c_2) .
\end{equation}
If there exists a positive integral $A_1$ of the \textsl{Abbott-curve} above the height level $c_1$ within 
the interval $[0, r_\mathrm{1}]$
\begin{equation}
A_1 \; = \; \int_0^{r_1} \, (c(r_\mathrm{M}) - c_1) \; \mathrm{d} \, r_\mathrm{M} \; > \; 0
\end{equation}
and for the dales a positive integral $A_2$
\begin{equation}
A_2 \; = \; \int_{r_2}^1 \, (c_2 - c(r_\mathrm{M})) \; \mathrm{d} \, r_\mathrm{M} \; > \; 0
\end{equation}
the parameters $R_\mathrm{pk}$ and $R_\mathrm{vk}$ are defined as
\begin{equation}
R_\mathrm{pk} \; = \; \frac{2 \, A_1}{r_1} \qquad \mathrm{and} \qquad 
R_\mathrm{vk} \; = \; \frac{2 \, A_2}{1 - r_2}.
\end{equation}
Topographies with amplitude distributions of kurtosis values much smaller than 3, for instance sinusoidal grids, have
no values above and below the core levels, i.e.\ no positive values for $A_1$ and $A_2$ and therefore no reduced peak
and dale heights.

We illustrate the effect of the choice of the \textsl{Abbott-curve} algorithm on the values of the material ratio parameters
using our measurements on ground steel. One of the profiles of scan length $L = 4 \, \mathrm{mm}$ and of a correlation length
$l_\mathrm{c} = 6.1 \, \mathrm{\mu m}$ being measured with a sampling interval of $\Delta x = 0.1 \, \mathrm{\mu m}$
is taken exemplarily.
The relative difference of the material ratio parameters whether obtained from the \textsl{Abbott-curve} by sorting
or by explicit material ratio evaluation lies below $10^{-4}$ if $\Delta x = 0.1 \, \mathrm{\mu m}$.
To reveal the effect, we reduced the resolution artificially by resampling the profile with an interval of
$\Delta x = 0.5 \, \mathrm{\mu m}$. To show the dependence on the raggedness we then evaluated the \textsl{Abbott-curve}
of the down sampled profile once without any software cut off of high frequencies, i.e.\ the lateral
limitation purely arises from the finite size of the probing sphere of a tip radius of approximately $2 \, \mathrm{\mu m}$.
In order to illustrate that the difference between the algorithms reduces the smoother the profile,
we simply performed some low pass filtration on the downsampled profile cutting off $\lambda_\mathrm{s} = 8 \, \mathrm{\mu m}$
and furthermore cutting off at a wavelength of $\lambda_\mathrm{s} = 25 \, \mathrm{\mu m}$. Regarding the difference
between the $R_\mathrm{k}$-value obtained by the sorting method $R_\mathrm{k, srt}$
and the $R_\mathrm{k}$-value obtained by material ratio calculation $R_\mathrm{k, mrc}$ interpolating linearly at the
intersection between height levels $c$ and asperity surfaces and the mean between those two values, we
evaluate following ratio to get the relative difference.
\begin{equation}
\Delta_\mathrm{rel}(R_\mathrm{k}) \; = \; \frac{R_\mathrm{k, srt} \, - \, R_\mathrm{k, mrc}}{\frac{1}{2} (R_\mathrm{k, srt} \, + \, R_\mathrm{k, mrc})}
\end{equation}
Evaluating the relative differences for all parameters delivers:

\begin{flushleft}
\begin{small}
\begin{tabular}{c||c|c|c}
\hline\hline
$\lambda_\mathrm{s}$/$\mathrm{\mu m}$  & -     & $8$      & $25$ \\
\hline\hline
$\Delta_\mathrm{rel}(R_\mathrm{k} )$  &  $8.7 \cdot 10^{-3}$ &   $1.1 \cdot 10^{-3}$ &  $0.5 \cdot 10^{-3}$ \\
$\Delta_\mathrm{rel}(R_\mathrm{pk})$ & $-7.0 \cdot 10^{-3}$ & $-15.7 \cdot 10^{-3}$ & $-6.0 \cdot 10^{-3}$ \\
$\Delta_\mathrm{rel}(R_\mathrm{vk})$ &  $5.6 \cdot 10^{-3}$ &   $5.0 \cdot 10^{-3}$ &  $0.4 \cdot 10^{-3}$ \\
\hline
$\Delta_\mathrm{rel}(r_1)$ & $-8.8 \cdot 10^{-3}$ & $4.2 \cdot 10^{-3}$ & $-6.0 \cdot 10^{-3}$\\
$\Delta_\mathrm{rel}(r_2)$ &  $1.5 \cdot 10^{-3}$ & $0.2 \cdot 10^{-3}$ & $< 10^{-4}$\\
\hline\hline
\end{tabular}
\end{small}
\end{flushleft}

\vspace{1mm}

The sorting approximation according to Eqn.~(\ref{zSort}) - (\ref{eqSortMr}) delivers the cumulative height
distribution of a topography. Therefore, the parameters $R_\mathrm{k}$, $R_\mathrm{pk}$, and $R_\mathrm{vk}$ are
directly related to the \textsl{PDF} of the height values. In the next two sections,
we will discuss the characteristics of \textsl{PDFs} of surface topographies in detail, first the way how sampling
influences its appearance and then we present the classification of \textsl{PDF} types in statistics.

\section{Influence of Sampling on ACF and PDF}
% Dokumente/ThomsenS-HRTS/stahl_rohling
% synthetic profiles with Fourier series
% invest_alias_pdf(100000, 100, 1.2, 3,-1); vs
% invest_alias_pdf(8000, 100, 1.2, 3,1);
% if (Delta_x/len_acf < 0.1 cut criterion as of OgilvyFoster1989) exp(-ln5*x'/ac_lenB) else exp(-ln5*(x'/ac_lenB).^2)
As topographies of rough surfaces still have regular structures, in particular those originating from machining
processes with rotating bodies thus producing periodic cutting traces, uniform sampling may cause aliasing and
leakage effects.
Therefore, the ratio between sampling interval $\Delta x$ and autocorrelation length $l_\mathrm{c}$ on one side
and the ratio between sampling length $L$ and autocorrelation length $l_\mathrm{c}$ on the other side are the determining
quantities for the reliability of the discretization of a topography. Consequently, Bushan suggests to use the correlation
length to define the sampling length $L$ \cite{BBushan2001}. Let $C\colon x \, \mapsto \, C(x)$ be the autocorrelation function
\begin{equation}
C(x) \; = \; \int_{-\infty}^\infty z(x) \, z(x - \xi) \, \mathrm{d} \xi
\end{equation}
and the autocorrelation length defined to be the length $x_\mathrm{c}$ where $C$ takes a certain value 
$C(x_\mathrm{c}) \, = \, C_\mathrm{c}$. Bushan sets $C_\mathrm{c} \, = \, 0.1$ denoting it $x_\mathrm{c} \, =: \, \beta^\ast$,
commonly $C_\mathrm{c} \, = \, e^{-1}$ as in \cite{OgilvyFoster1989} denoting it $x_\mathrm{c} \, =: \, \lambda_0$,
and in ISO 25178-2 it is $C_\mathrm{c} \, = \, 0.2$. In this article, we define
$x_\mathrm{c} \, =: \, l_\mathrm{c}$ for $C_\mathrm{c} \, = \, 0.2$. Bushan's suggestion of an appropriate profile
length of random surfaces to be
\begin{equation}
L \; \ge \; 200 \, \beta^\ast
\end{equation}
%For an exponential autocorrelation function it is 
%\begin{equation}
%L \; \ge \; 200 \, \frac{\mathrm{ln}(0.1)}{\mathrm{ln}(0.2)} \, l_\mathrm{c} \; \approx \; 286 \, l_\mathrm{c}
%\end{equation}
%for a Gaussian
%\begin{equation}
%L \; \ge \; 200 \, \sqrt{\frac{\mathrm{ln}(0.1)}{\mathrm{ln}(0.2)}} \, l_\mathrm{c} \; \approx \; 239 \, l_\mathrm{c}
%\end{equation}
means that $L$ should be around $300 \, l_\mathrm{c}$. We have examined a ground steel surface with
$300 \, l_\mathrm{c} \, = \, 1.6 \dots 2.4 \, \mathrm{mm}$ which is about half of the sampling length according to ISO 4288 of
$4 \, \mathrm{mm}$. We also have examined a ceramics surface of cutting tool inserts with
$300 \, l_\mathrm{c} \, = \, 1.8 \dots 3.0 \, \mathrm{mm}$ and a few outlying profiles, where
$300 \, l_\mathrm{c}$ took values above the  $L = 4 \, \mathrm{mm}$ in the range of $4.3 \dots 5.7 \, \mathrm{mm}$.

For the sampling interval Bushan suggests $\Delta x \; < \; 0.25 \, \beta^\ast$, i.e.\
$\Delta x \; < \; 0.35 \, l_\mathrm{c}$, at least $\Delta x \; < \; 0.5 \, \beta^\ast$. 
% /home/hueser01/Dokumente/1-Algos_devel/Synthetic_Rough_for_MC/OgilvyFoster-Rough_surfaces-gaussian_or_exponential_statistics-1989.pdf
In 1989, Ogilvy and Foster \cite{OgilvyFoster1989} have examined the influence of the sampling interval on the shape of the
resultant autocorrelation function and its deviation from the original exponential progress. They state that 
a sampling interval of $\Delta x \; < \; 15^{-1} \, \lambda_0$ ($\Delta x \; < \; 0.04 \, l_\mathrm{c}$) would be adaequate to
detect the exponential nature of the autocorrelation function, which according to them is most likely for rough surface
topographies, thus for the surfaces under investigation at around
$\Delta x \, = \, 0.2 \dots 0.3 \, \mathrm{\mu m}$. According to ISO standard our surfaces should be sampled
with at most $\Delta x \, = \, 0.5 \, \mathrm{\mu m}$ and we have measured with a sampling
interval of $\Delta x \, = \, 0.1 \, \mathrm{\mu m}$. The suggestions of Bushan originate from the late 1980s and
beginning of 1990s, while nowadays instrumental and computational technologies allow broader bandwidths.

Finite and uniform sampling causes an exponential autocorrelation of a rough surface to show ripples like a sinc function or a
Bessel function, since they are caused by the convolution with an impulse train of Dirac or box pulses. In order to illustrate
the relation between sampling and the shape of the \textsl{ACF} as well as the shape of the \textsl{PDF}, we have generated
roughness profiles that we could describe analytically choosing Fourier series of a finite set of spatial frequencies
that means of reciprocal wavelengths $\lambda$. Two types of
probability density distributions of the frequency sets are compared: a one-sided Gaussian and a uniform distribution.
For the one-sided Gaussian we employ
\begin{equation}
\mathcal{N}(\lambda_\mathrm{m}, \lambda_\mathrm{BW}) \; \propto \;
e^{-\frac{1}{2} \, \left(\frac{\lambda \, - \, \lambda_\mathrm{m}}{\lambda_\mathrm{BW}}\right)^2 } \qquad 
\lambda \, \ge \, \lambda_\mathrm{m}
\end{equation}
where $\lambda_\mathrm{BW}$ denotes the width and $\lambda_\mathrm{m}$ denotes the center
of the Gaussian distribution and the maximum probability. With $\lambda \, \ge \, \lambda_\mathrm{m}$ the parameter
$\lambda_\mathrm{m}$ denotes the left border of the interval so the smallest wavelength (highest spatial frequency).
The diced wavelengths $\lambda$ will scatter close to $\lambda_\mathrm{m}$ on the right side,
i.e.\ within an interval of about $[\lambda_\mathrm{m}, \lambda_\mathrm{m}+3 \lambda_\mathrm{BW}]$.
%an exponential
%\begin{equation}
%\mathcal{E}(\lambda_\mathrm{m}, \lambda_\mathrm{BW}) \; \propto \;
%e^{-\frac{\lambda \, - \, \lambda_\mathrm{m}}{\lambda_\mathrm{BW}} } \qquad 
%\lambda \, \ge \, \lambda_\mathrm{m}
%\end{equation}
For the uniform distribution we use
\begin{equation}
\mathcal{U}(\lambda_\mathrm{m}, \lambda_\mathrm{BW}) \; \propto \;
\left\{\begin{array}{cl}
const. & \lambda \, \in \, [\lambda_\mathrm{m}, \lambda_\mathrm{m}+\lambda_\mathrm{BW}] \\
0 & \mathrm{else}
\end{array}\right. .
\end{equation}
where $\lambda_\mathrm{BW}$ denotes the width of the scattering interval and $\lambda_\mathrm{m}$ the left side of the
interval.

Sets of $N$ wavelengths $\{\lambda_1, \dots, \lambda_N\}$ are diced according to the above listed
distributions, furthermore $N$ phases $\{\varphi_1, \dots, \varphi_N\}$ are diced according to a uniform distribution with
$\varphi_\nu \, \in \, [-\pi, \pi]$, and amplitudes $a_\nu \, \propto \, \mathrm{exp}(-\lambda_\nu / (10^{-3} \, \mathrm{\mu m}))$
were chosen, $\nu = 1, \dots, N$. A continuous profile is synthesized for $x$ being the continuous lateral position, i.e.\ for 
$x \in \mathrm{I \! R}$
\begin{equation}
z(x) \; = \; \sum_{\nu = 1}^N a_\nu \, \sin \left(\frac{2 \, \pi}{\lambda_\nu} \, x \; + \; \varphi_\nu \right) .
\end{equation}
For a Dirac pulse sampling the profile is discretized as
\begin{equation}
z(x_i) \; = \; \sum_{\nu = 1}^N a_\nu \, \sin \left(\frac{2 \, \pi}{\lambda_\nu} \, (i-1) \, \Delta x \; + \; \varphi_\nu \right)
\end{equation}
with $i = 1, \dots, n$ and for a box pulse train with a pulse width $w$ the profile samples are
\begin{equation}
z(x_i) \; = \; \sum_{\nu = 1}^N a_\nu \, \int_{A_i}^{B_i} 
\sin \left(\frac{2 \, \pi}{\lambda_\nu} \, \xi \; + \; \varphi_\nu \right) \,  \mathrm{d} \xi
\end{equation}
with 
$$
A_i \; = \; (i-1) \, \Delta x \, - \, \frac{w}{2} ;
\qquad
B_i \; = \; (i-1) \, \Delta x \, + \, \frac{w}{2} .
$$
% 
% ac_len5 =  8.6001
% ac_len7 =  11.560

\begin{figure}
\begin{center}
\includegraphics[width=75mm,keepaspectratio]{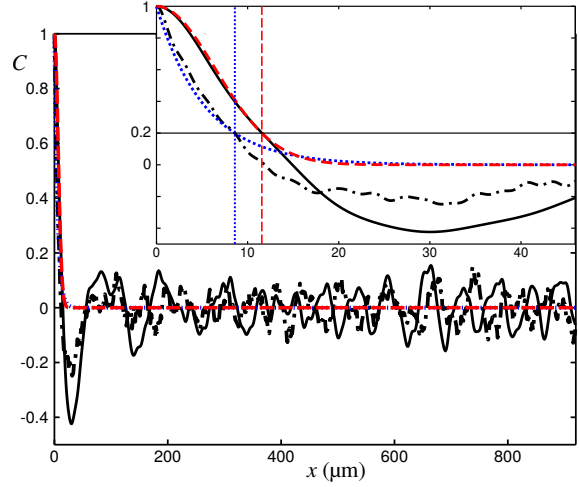}
\end{center}
\caption{Autocorrelation functions of two different Fourier series: 1. from uniformly distributed spatial wavelengths 
(black solid curve) with $\lambda_\mathrm{m} \, = \, 11.3 \, \mathrm{\mu m}$
and $\lambda_\mathrm{BW} \, = \, 102 \, \mathrm{\mu m}$, correlation length of $l_\mathrm{c} \, = \, 11.56 \, \mathrm{\mu m}$
and Gaussian \textsl{ACF} (red dashed curve); 2. from Gaussian distributed wavelengths (black dash-dotted curve) with
$\lambda_\mathrm{m} \, = \, 2.6 \, \mathrm{\mu m}$ and $\lambda_\mathrm{BW} \, = \, 53 \, \mathrm{\mu m}$,
correlation length of $l_\mathrm{c} \, = \, 8.6 \, \mathrm{\mu m}$ and exponential \textsl{ACF} (blue dotted curve).}
\label{fig:ACFGaussianvsExp1}
\end{figure}

\begin{figure}
\begin{center}
\includegraphics[width=75mm,keepaspectratio]{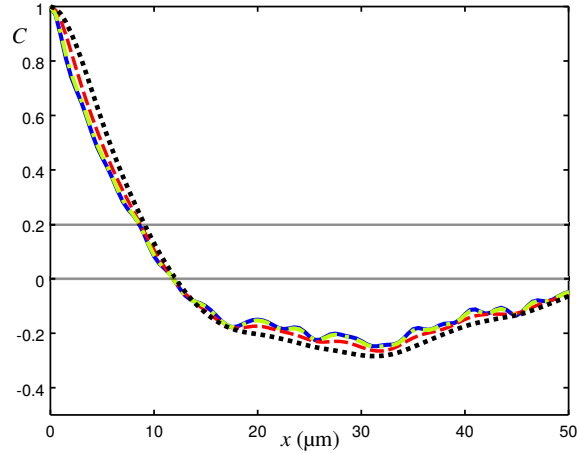}
\end{center}
\caption{Autocorrelation functions of the Fourier series after being sampled differently:
1. Dirac impulse train (blue solid curve),
2. box impulse train with $w = 0.8 \, \mathrm{\mu m}$ (light green dash-dotted curve), 
3. box impulse train with $w = 2.5 \, \mathrm{\mu m}$ (red dashed curve),
4. box impulse train with $w = 5.0 \, \mathrm{\mu m}$ (black dotted curve).}
\label{fig:ACFGaussianvsExp2}
\end{figure}

The origin of ripples of the autocorrelation function $C$ may as well be due to the finite sample size of contributing spatial
frequencies $N$. Furthermore, the kind of distribution of the frequencies contributing to the Fourier series determines
whether $C$ is better represented by an exponential or by a Gaussian function. Fig.~\ref{fig:ACFGaussianvsExp1} shows
the \textsl{ACFs} of two different Fourier series, both with $N = 431$ spatial frequencies. The black solid curve is the
one obtained from the profile with uniformly distributed spatial wavelengths with
$\lambda_\mathrm{m} \, = \, 11.3 \, \mathrm{\mu m}$
and $\lambda_\mathrm{BW} \, = \, 102 \, \mathrm{\mu m}$ delivering a correlation length of
$l_\mathrm{c} \, = \, 11.56 \, \mathrm{\mu m}$, which is to be compared to the Gaussian \textsl{ACF} visualized as red dashed curve.
%	N = 431;
%#	la = 53*rande(N,1); file numbers 4
%#	la = 53*randn(N,1); file numbers 5
%#	la = 53*rand(N,1); #file numbers 6
%#	las = sort(abs(la),"descend") + 2.6; #file numbers 4,5,6
%#	la = 102*rand(N,1); #file numbers 7
%#	las = sort(abs(la),"descend") + 11.3; #file number 7
The black dash-dotted curve is the \textsl{ACF} obtained from the profile with Gaussian distributed wavelengths with
$\lambda_\mathrm{m} \, = \, 2.6 \, \mathrm{\mu m}$
and $\lambda_\mathrm{BW} \, = \, 53 \, \mathrm{\mu m}$ delivering a correlation length of
$l_\mathrm{c} \, = \, 8.6 \, \mathrm{\mu m}$, to be compared to the exponential \textsl{ACF} visualized as blue dotted curve
and to which we will refer as \textsl{profile G}.
In Fig.~\ref{fig:ACFGaussianvsExp1} we can see that significant high frequency contributions owe the exponential behavior of
the \textsl{ACF}. Be it due to lower resolution as investigated by Ogilvy and Foster or due to the fact that there exist as
little high frequencies as low frequencies as we have calculated it here. The resultant \textsl{ACFs} have a Gaussian behavior
in both cases.
To show the effect of lateral resolution, we have calculated the discretization of \textsl{profile G}
for a Dirac impulse train, and box impulse trains with three different widths $w$.
Fig.~\ref{fig:ACFGaussianvsExp2} shows the \textsl{ACF} of the data by Dirac impulse train as blue solid curve,
those of box impulse trains with
$w = 0.8 \, \mathrm{\mu m}$ as light green dash-dotted curve, $w = 2.5 \, \mathrm{\mu m}$ as red dashed curve, and
$w = 5.0 \, \mathrm{\mu m}$ as black dotted curve. Since the Fourier series minimum value of wavelength is 
$\lambda_\mathrm{m} = 2.6 \, \mathrm{\mu m}$, the difference between the box (light green dash-dotted) and the Dirac
impulse (blue solid) trains is negligible. In accordance with Ogilvy and Foster the larger the box width, i.e.\
the smaller the resolution, the more the \textsl{ACF} turns to a Gaussian curve shape.

\begin{figure}
\begin{center}
\includegraphics[width=75mm,keepaspectratio]{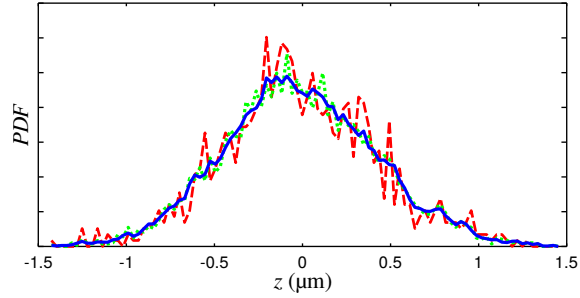}
\end{center}
\caption{Probability density distribution $\Delta x = 0.05 \, \mathrm{\mu m}$ solid blue curve,
 $\Delta x = 0.8 \, \mathrm{\mu m}$ dotted green, $\Delta x = 4 \, \mathrm{\mu m}$ dashed red curve.}
\label{fig:PDFresolutionDx}
\end{figure}

\begin{figure}
\begin{center}
\includegraphics[width=75mm,keepaspectratio]{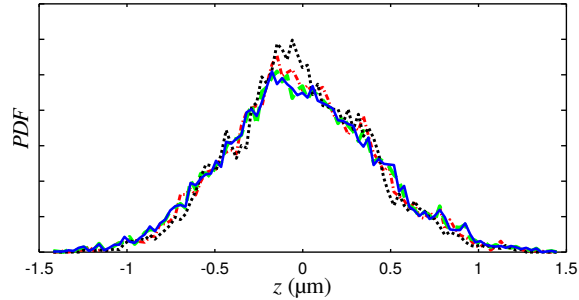}
\end{center}
\caption{Probability density distribution Dirac solid blue curve, $w = 0.8 \, \mathrm{\mu m}$ dashed green curve,
 $w = 2.5 \, \mathrm{\mu m}$ dash-dotted red, $w = 5 \, \mathrm{\mu m}$ dotted black curve.}
\label{fig:PDFresolutionBox}
\end{figure}

As the \textsl{ACF} it is also the \textsl{probability density distribution PDF}, which is affected by the resolution
issue. Fig.~\ref{fig:PDFresolutionDx} shows the \textsl{PDFs} of \textsl{profile G} for three different sampling intervals,
with all of them sampled with Dirac pulses. The length of the profile has always been $4 \, \mathrm{mm}$ such that the
sample size (number of points) $n$ decreases: $\Delta x = 0.05 \, \mathrm{\mu m}$ with $n = 80 \, 000$ is
plotted as solid blue curve, $\Delta x = 0.8 \, \mathrm{\mu m}$ with $n = 5000$ as dotted green curve, and
$\Delta x = 4 \, \mathrm{\mu m}$ with $n = 1000$ as dashed red curve.
% invest_alias_pdf_2(nbins) nbins=100
The \textsl{PDF} is pronged due to the smaller sample size. We have also investigated the effect of the impulse box
size on the \textsl{PDFs} for fixed sample sizes.
One example for sample size $n = 1000$ and different impulse trains is shown in Fig.~\ref{fig:PDFresolutionBox}.
The blue curve shows the \textsl{PDF} of a profile sampled by a Dirac impulse train,
the dashed green curve the \textsl{PDF} with box pulse sampling of width $w = 0.8 \, \mathrm{\mu m}$,
the red dash-dotted curve with $w = 2.5 \, \mathrm{\mu m}$, and the black dotted curve with $w = 5 \, \mathrm{\mu m}$.

\section{PDF Types Parameterizing Roughness Amplitude Distributions}

During the past decades, a variety of investigations were made to classify surfaces obtained from different kind of engineering
processes according to their amplitude distributions and how they deviate from Gaussian distribution according to
their statistical moments and furthermore according to anisotropy and lay.

In 1994, Whitehouse \cite{Whitehouse1994ch2} suggested to use different classes of $\beta$-distributions.
He derived a relation between the parameters of the $\beta$-distributions and the statistical moments. To describe data
that are distributed with long tails generalized extreme value functions are employed, in particular in the field of
finance statistics \cite{Gilli2006}. Extreme value theory has been developed to characterize data with values that extremely
deviate from the \textsl{median} of probability distributions \cite{Gilli2006}. The \textsl{median} is a robust
statistical parameter defined to be the middle value of the sorted set of discrete observations of a quantity.
%Another robust estimator of the \textsl{PDF} is the \textsl{arithmetic average roughness} being defined without mean
%subtraction in roughness metrology as
%\begin{equation}
%R_\mathrm{a} \; = \; \frac{1}{L} \int_0^L \mid z (x) \mid \mathrm{d} x .
%\end{equation}

% Dokumente/pubs_literat_all/pubs_reports/25_RkUncertainty/lit/Gilli_An_application_of_EVT_for_measuring_financial_risk-2006.pdf
Weibull distributions, as an example of an extreme value distribution, are of particular interest for
honed cylinder liner or cylinder running surfaces with oil volume and the elastic-plastic contact of
asperities \cite{YuPolycarpou2002}.
% Dokumente/pubs_literat_all/pubs_reports/25_RkUncertainty/lit/YuPolycarpou-Contact_of_rough_surfaces_with_asymmetric_distribution_of_asperity_heights__Weilbull-2002.pdf

Since the functions of the Johnson system define non-Gaussian probability density functions that 
facilitate simulations based on white noise by explicit and invertable transformation
$f\colon z \mapsto t(z)$ \cite{Johnson1949}. We employ Johnson distributions
for the Monte Carlo simulation of profiles. They are represented by
Gaussian distributions of a transformation of the quantity to be examined, i.e.\
$\propto exp(-0.5 \, t^2)$. Their relation to statistical moments can be estimated by the Hill \textsl{et al}
optimization algorithm of 1976 \cite{HillHillHolder1976}, which is available as Matlab/Gnu-octave
routine as well. Johnson defines following three types of transformation functions $f$:
\begin{itemize}
\item the lognormal system \textsl{SL}
\begin{equation}
t \; = \; \gamma \, + \, \delta \, \mathrm{ln} \left(z - \xi \right) \qquad \xi < z
\end{equation}
\item the unbounded system \textsl{SU}
\begin{equation}
t \; = \; \gamma \, + \, \delta \, \mathrm{arcsinh} \left(\frac{z - \xi}{\lambda} \right)
\end{equation}
\item the bounded system \textsl{SB}
\begin{equation}
t \; = \; \gamma \, + \, \delta \, \mathrm{ln} \left(\frac{z - \xi}{z + \lambda - \xi} \right) \; \; \xi < z < \xi + \lambda
\end{equation}
\end{itemize}
Normally distributed random numbers of a quantity $t$ can then be transformed by using the inverse transformation
$f^{-1} \colon t \mapsto z(t)$.

% /home/hueser01/Dokumente/1-Algos_devel/Synthetic_Rough_for_MC/Ashworth_etal-Representation of ion implantation profiles by Pearson frequency distribution curves-1990-0022-3727_23_7_018.pdf
% /home/hueser01/Dokumente/ThomsenS-HRTS/stahl_rohling/Zhuang-Gaussian_mixture_density_model-00535841.pdf
% /home/hueser01/Dokumente/1-Algos_devel/Synthetic_Rough_for_MC/ChilamakuriBhushan-Contact_analysis_of_non-Gaussian_random_surfaces-1998.pdf

% family of alternative distributions:
% /home/hueser01/Dokumente/pubs_literat_all/pubs_reports/25_RkUncertainty/lit/HoffmanHammonds-Propagation_of_uncertainty_inrisk_assessment-1994.pdf
To use the algorithm of Hill \textsl{et al}, we employ the statistics definition of
the statistical moments of the \textsl{PDF}, i.e.\ those with mean subtraction, whereas roughness standards
ISO 4287 and ISO 25178-2 define these statistical moments without mean subtraction presuming
that the detrending of waviness by cutting off spatial frequencies below $\lambda_\mathrm{c}^{-1}$
causes the mean $\bar z$ to be very close to zero and hence negligible, i.e.\

\begin{equation}
\bar z \; = \;  \frac{1}{L} \int_0^L z (x) \mathrm{d} x  \; \approx \; 0 .
\label{moment1}
\end{equation}

The third moment in roughness metrology is
\begin{equation}
R_\mathrm{sk} \; = \; \frac{1}{L \, R_\mathrm{q}^3} \int_0^L \left(z (x)\right)^3 \mathrm{d} x
\end{equation}
while in statistics and as input for Hill \textsl{et al} we use
\begin{equation}
\mu_3 \; = \; \frac{1}{n \, s^3} \sum_{i=1}^n \left(z_i - \bar z \right)^3
\label{moment3}
\end{equation}
with $\bar z$ being the mean of all $z_i$, called first moment.
The fourth moment or kurtosis is
\begin{equation}
R_\mathrm{ku} \; = \; \frac{1}{L \, R_\mathrm{q}^4} \int_0^L \left(z (x)\right)^4 \mathrm{d} x
\end{equation}
respectively
\begin{equation}
\mu_4 \; = \; \frac{1}{n \, s^4} \sum_{i=1}^n \left(z_i - \bar z \right)^4
\label{moment4}
\end{equation}
with the second moment being
\begin{equation}
R_\mathrm{q}^2 \; = \;  \frac{1}{L} \int_0^L \left(z (x)\right)^2 \mathrm{d} x 
\end{equation}
respectively the variance, i.e.\ the square of the standard deviation $s$
\begin{equation}
s^2 \; = \; \mu_2 \; = \; \frac{1}{n - 1} \sum_{i=1}^n \left(z_i - \bar z \right)^2 .
\label{moment2}
\end{equation}
and with $x$ being the lateral position, $L$ the length of the profile, $n$ the number of sampling points, and
$z$ the roughness profile after band limitation by detrending filtration.

% Dokumente/1-Algos_devel/Synthetic_Rough_for_MC/ChilamakuriBhushan-Contact_analysis_of_non-Gaussian_random_surfaces-1998.pdf
% Dokumente/1-Algos_devel/Synthetic_Rough_for_MC/Bakolas-Numerical_generation_of_arbitrarily_oriented_non-Gaussian_3D_rough_surfaces-2003.pdf

% Dokumente/1-Algos_devel/bosse_lit_rough_stepheight/Lopes_etal-A_multivariate_surface_roughness_modeling_and_optimization-2013.pdf

\begin{figure}
\begin{center}
\includegraphics[width=75mm,keepaspectratio]{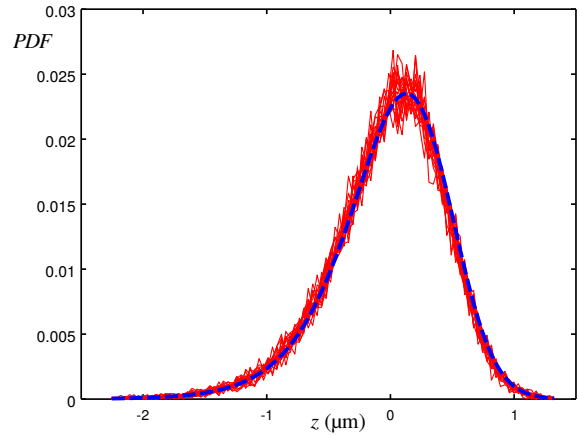}
\end{center}
\caption{Probability density functions (red solid curves) obtained by simulating $20$ profiles according to its
autocorrelation length, which has a value of $l_\mathrm{c} = 6 \, \mathrm{\mu m}$, and with $\Delta x = 0.4 \, \mathrm{\mu m}$
using the Johnson's system function (blue dashed curve).}
\label{fig:PDFsteel999simuLR}
\end{figure}

Now we regard our measurement on ground steel with $\Delta x = 0.1 \, \mathrm{\mu m}$ and $L = 4 \; \mathrm{mm}$,
i.e.\ with $n = 40 \, 000$ with scanning direction orthogonal to the lay.
% Dokumente/ThomsenS-HRTS/stahl_rohling
% invest_meas_pdf(100)
\begin{figure}
\begin{center}
\includegraphics[width=75mm,keepaspectratio]{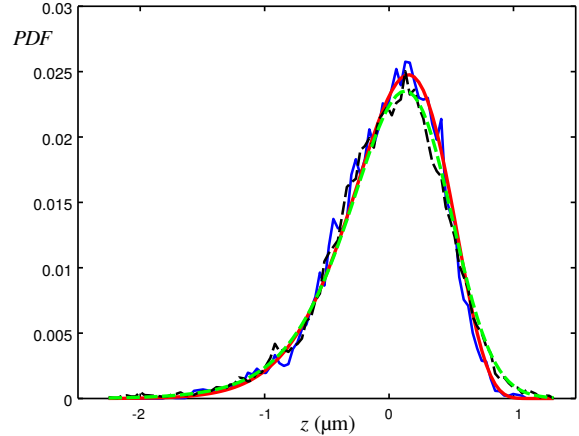}
\end{center}
\caption{Probability density distributions of 2 scans on a ground steel surface
 sanned perpendicular to the lay with $\Delta x = 0.1 \, \mathrm{\mu m}$ and their
 estimates of Johnson system's functions: the \textsl{PDF} of the data plotted
as black dashed curve is parameterized with a function of the unbounded system (green dashed curve);
the \textsl{PDF} plotted as blue solid curve with a function of the bounded system (red solid curve).}
\label{fig:PDFsteel000and999}
\end{figure}
Estimation of a \textsl{PDF} of the Johnson system via statistical moments delivered functions of the \textsl{SU} type
for some of the profiles and the \textsl{SB} type for most of the profiles. Fig.~\ref{fig:PDFsteel000and999} shows
exemplarily two of $1000$ parallel profiles. The two displayed profiles lie $4 \; \mathrm{mm}$ appart,
one parameterized with a \textsl{PDF} of the \textsl{SU} the other of the \textsl{SB} type.
Using the Johnson's system function (green dashed curve) of the second profile's \textsl{PDF} plotted as black dashed
curve and simulating $20$ profiles according to its autocorrelation length, which has a value of
$l_\mathrm{c} = 6 \, \mathrm{\mu m}$, and with $\Delta x = 0.4 \, \mathrm{\mu m}$ delivers
the \textsl{PDFs} shown in Fig.~\ref{fig:PDFsteel999simuLR}.
% invest_meas_vs_john_pdf_LR(100)

\begin{figure}
\begin{center}
\includegraphics[width=75mm,keepaspectratio]{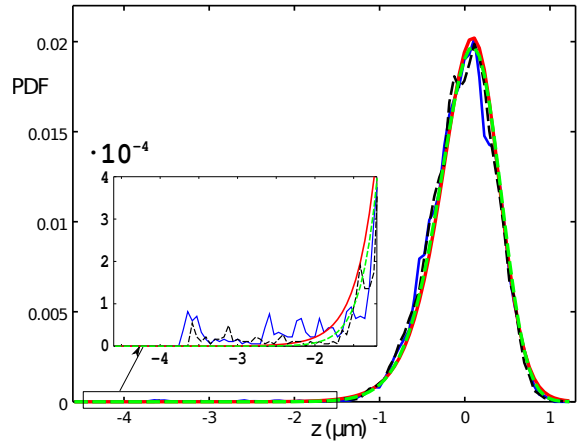}
\end{center}
\caption{Probability density distribution of 2 example scans on a ceramics surface. The measurment
 plotted as solid blue curve has been parameterized with a Johnson's \textsl{SU} function, which
 is drawn as solid red curve; the measurment drawn as dashed black curve estimated with that
 of dashed green curve. Moment estimations required a cut off of values $z \le -2.5 \, \mathrm{\mu m}$ to deal
 with the extraoridinarily long tails to reproduce the shape of the core of the \textsl{PDFs}.}
\label{fig:PDFceramic000and500}
\end{figure}

% Dokumente/ThomsenS-HRTS/keramik_schneid
% invest_meas_pdf_keramik(100) gt -2.5 um
The ceramics sample that we have investigated shows extremely deep pores such that the \textsl{PDFs} have a
significant long left tail biasing the statistical moments. To obtain statistical moments delivering appropriate
Johnson system functions, we have eliminated the height values below $-2.5 \, \mathrm{\mu m}$ when evaluating
the statistical moments. Fig.~\ref{fig:PDFceramic000and500} shows the \textsl{PDFs} of two of the measured profiles
together with the estimated \textsl{SU} functions for each of them.
The requirement of tail elimination for the scans on the ceramics surface shows the limits of the
procedure to employ statistical moments for a subsequent estimation of Johnson system's
functions. For a more intricate investigation, hence more complex Monte Carlo approaches,
a mixing of more than one stochastic process is needed. A mixing may be achieved by superimposing
profiles and finally \textsl{PDFs}. Typical examples for such superpositions are surfaces obtained by different steps
of machining processes such as honing after grinding. The grinding delivers deep grooves and dales, for instance
as oil volume, and the honing smoothes the upper part of the surface to decrease friction. Pawlus superimposes
two Gaussian distributed surfaces to simulate this type of surfaces \cite{Pawlus2008}.
% pubs_literat_all/Literatur/Rauheit_TechnischeOberfl/Rauheitsstochastik/Pawlus-Simu_of_stratified_surface_topos-2007.pdf
This approach will be investigated more thoroughly with regard to the statistics of topographies
produced by different machining steps. We are considering a combination of Gaussian and non-Gaussian \textsl{PDFs}
for future work \cite{Rief201x}.
% Rief Diss to be published
A generalized approach of a superpositioning of Gaussian processes is the
Gaussian mixture density modeling, which has already been investigated twenty years ago \cite{ZhuangHuang1996}.
% /home/hueser01/Dokumente/ThomsenS-HRTS/stahl_rohling/Zhuang-Gaussian_mixture_density_model-00535841.pdf

\section{Approximating Material Ratio Uncertainties}

In this section, we now present a procedure for estimating the uncertainty contribution to roughness parameters
caused by the stochasticity of the topography of the measured surface. In order to
investigate its possibilities and limitations we have examined surfaces of different material and different type of
machining, topographies with lay and isotropical textures. All surfaces under investigation for this article have
$R_{\mathrm{k}}$ values in the order of magnitude of $1 \; \mathrm{\mu m}$. They show standard deviations of
$R_{\mathrm{k}}$ that lie in the range of two to five percent.
The standard deviation of $R_{\mathrm{pk}}$ and that of $R_{\mathrm{vk}}$ are in the same order of
magnitude than that of $R_{\mathrm{k}}$ in their absolute values, such that their relative deviation is greater accordingly.
To demonstrate our Monte Carlo approach with its advantages and drawbacks here,
we have chosen the profiles measured on our ground steel sample.

The proposed procedure only provides for an \textsl{approximate} estimate of the uncertainty of material ratio parameters.
It does not replace the characterization of the texture over a macroscopic range.
The method procedes as follows:
\begin{enumerate}
\item numerical evaluation of \textsl{ACF} of the discrete profile of $n$ height values $\{z_1, \dots , z_n\}$
 with calculation of the autocorrelation length $l_\mathrm{c}$ by selecting the first intersection
 of the \textsl{ACF} curve with $C = 0.2$;
\item choose the appropriate model for \textsl{ACF}, either exponential or Gaussian, then
 evaluate $C(x,l_\mathrm{c})$ and its power spectrum density for a discrete sample of sample size $n$;
 use Fourier transform of weights $\{w_1, \dots , w_n\}$ that are derived from $C(x_i,l_\mathrm{c})$
 according to \cite{OgilvyFoster1989};
% (either upsampled $n = \nu \, m$ with $\nu \in \mathrm{I} \! \mathrm{N}$ (a positive interger)
% or same sample size $n = m$ according to the appropriateness of $\Delta x$ of measured profile);
\item estimate the statistical moments of the profile $\{z_1, \dots , z_n\}$ according to
 Eqn.~(\ref{moment1}), (\ref{moment3}), (\ref{moment4}), and (\ref{moment2}) and estimate function
 of Johnson system;
\item perform $K$ times (for instance $K = 100$), i.e.\ $\kappa = 1, \dots,K$ the following sub-steps:
 \begin{enumerate}
  \item generate random white noise $\{r_1, \dots , r_n\}_\kappa$ of sample size $n$ and convolve it with
   the weights $\{w_1, \dots , w_n\}$ such that a correlated sequence of values $\{t_1, \dots , t_n\}_\kappa$
    \begin{equation}
   t_i = \sum_\nu w_{\nu} r_{i + \nu}
   \end{equation}
   is obtained with $w_\nu$ such that
   \begin{equation}
   C(x_i,l_\mathrm{c}) = \sum_\nu t_{\nu} t_{i + \nu}
   \end{equation}
   by multiplication in Fourier space accordingly;
  \item transform sequence $\{t_1, \dots , t_n\}_\kappa$ via inverse function $f^{-1}$ of Johnson system's function
%   (if $m < n$ downsample it appropriately)
   to obtain a simulated profile
   $\{{\tilde z}_1, \dots , {\tilde z}_n\}_\kappa$
  \item evaluate the set of material ratio parameters $R_{\mathrm{k}, \kappa}$, $R_{\mathrm{pk}, \kappa}$, etc.\ 
   of the simulated profile $\{{\tilde z}_1, \dots , {\tilde z}_n\}_\kappa$
 \end{enumerate}
\item evaluate mean and standard deviation of each of the material parameters over $K$ values:
$$
\bar R_{\mathrm{k}} \; = \; \frac{1}{K} \sum_{\kappa = 1}^K R_{\mathrm{k}, \kappa} \qquad \mathrm{etc.}
$$
and
$$
s (R_{\mathrm{k}}) \; = \; \sqrt{\frac{1}{K-1} \sum_{\kappa = 1}^K 
  \left( R_{\mathrm{k}, \kappa} \, - \, \bar R_{\mathrm{k}} \right)^2 } \qquad \mathrm{etc.}
$$
\end{enumerate}
%For an estimation of the standard deviations valid for a macroscopic range of a surface, non-uniformly,
%randomly distributed profiles are to be taken and an analysis of the statistical moments of those profiles
%has to be carried out. Then a model is needed, at least some simple functional relation like a linear or
%quadratic between skewness and kurtosis may be estimated in order to generate random variations of the
%moments as well and include step (iii) into the loop of step (iv).

\begin{figure}
\begin{center}
\includegraphics[width=70mm,keepaspectratio]{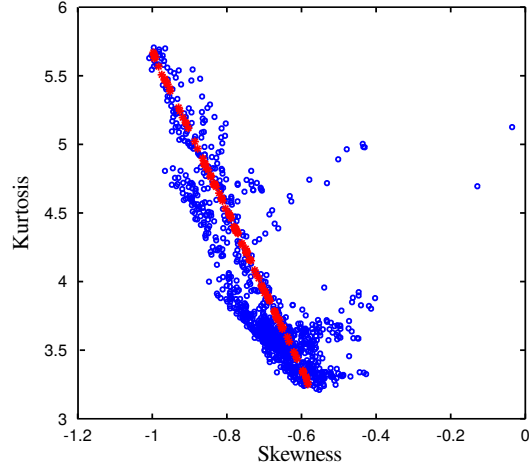}
\end{center}
\caption{Relation between skewness and kurtosis for a ground steel surface, the blue circles are
obtained by experiment and the red asterix are values used for Monte Carlo simulation to be passed to
the Johnson system of \textsl{PDFs}.}
\label{fig:KurtvsSkew}
\end{figure}

On ground steel, we have measured $1000$ profiles with a sampling interval of $\Delta x = 0.1 \, \mathrm{\mu m}$.
The mean observed autocorrelation length is $l_\mathrm{c} = 6 \, \mathrm{\mu m}$.
The covered area takes $4 \, \times \,4 \, \mathrm{mm}^2$. We have observed significantly varying $R_\mathrm{q}$ values as
well as varying skewness and kurtosis as can be seen in Fig.~\ref{fig:KurtvsSkew} as blue circles.

\vspace{3mm}

\begin{tabular}{l||c|c|c|c|c}
\hline\hline
\multicolumn{6}{l}{\textbf{Empirical result of $1000$ measured profiles}} \\
\hline\hline
 & $R_\mathrm{k}$ & $R_\mathrm{pk}$ & $R_\mathrm{vk}$ & Mr1 & Mr2 \\
 & $\mathrm{\mu m}$ & $\mathrm{\mu m}$ & $\mathrm{\mu m}$ & \% & \% \\
\hline\hline
mean  & 1.07 & 0.27 & 0.64 &   7.2 & 86.3 \\
$s$ & 0.04 & 0.05 & 0.04 &   0.8 &  0.9 \\
\hline\hline
\end{tabular}
\vspace{3mm}

The standard deviations $s$ of $R_\mathrm{k}$, $R_\mathrm{pk}$, and $R_\mathrm{vk}$ have values of
around $50$ Nanometer. To compare the experimental result with our Monte Carlo method, we
have run the Monte Carlo simulation $20$ times with $100$ profiles each, with
$L = 4 \, \mathrm{mm}$ and $l_\mathrm{c} = 6 \, \mathrm{\mu m}$. To show the influence of resolution 
we have done this for three different sampling intervals
$\Delta x = 0.1 \, \mathrm{\mu m}$, $\Delta x = 0.5 \, \mathrm{\mu m}$, and $\Delta x = 1 \, \mathrm{\mu m}$, 
delivering following values for the standard deviation of each of the material ratio parameters:
\begin{flushleft}
\begin{tabular}{c||c|c|c}
\hline\hline
\multicolumn{4}{l}{\textbf{Influence of resolution:}} \\
\multicolumn{4}{l}{\textbf{Standard dev.\ of $20$ simulations of $100$ profiles}} \\
\hline\hline
$\Delta x$/$\mathrm{\mu m}$            & $0.1$      & $0.5$      & $1$ \\
\hline\hline
$s(R_\mathrm{k})$/$\mathrm{n m}$  & $23 \pm 2$ & $25 \pm 2$ & $28 \pm 2$ \\
$s(R_\mathrm{pk})$/$\mathrm{n m}$ & $15 \pm 1$ & $16 \pm 1$ & $17 \pm 2$ \\
$s(R_\mathrm{vk})$/$\mathrm{n m}$ & $30 \pm 2$ & $30 \pm 2$ & $32 \pm 2$ \\
\hline
$s$(\small{Mr1})/ \%  & $0.48 \pm 0.03$ & $0.53 \pm 0.03$ & $0.60 \pm 0.05$\\
$s$(\small{Mr2})/ \%  & $0.69 \pm 0.06$ & $0.74 \pm 0.06$ & $0.78 \pm 0.06$\\
\hline\hline
\end{tabular}
\end{flushleft}

\vspace{1mm}

The values given in the above table show that the uncertainty of the core height lies around three percent,
the uncertainty of the reduced peak and dale heights around six percent.

Similar to the distribution of skewness and kurtosis of the experimental data we have randomly changed the
moments for the Johnson functions within the Monte Carlo loop (included step (iii) into step (iv)). Here, we show
one example with which we mimic the experimental relation of skewness and kurtosis of the ground surface shown in Fig.~\ref{fig:KurtvsSkew}
as blue circles. We have varied the skewness $\mu_3$ due to a uniform distribution then evaluated the kurtosis $\mu_4$
to be close to following straight line segment:
$$
\left(\begin{array}{c}
\mu_3\\
\mu_4
\end{array}\right)
\in
\left[
\left(\begin{array}{c}
-1.0\\
5.67
\end{array}\right), \,
\left(\begin{array}{c}
-0.58\\
3.25
\end{array}\right)
\right]
$$
The kurtosis has been normally distributed around that line with $\sigma_{\mu_4} = 0.01$. The diced pairs
of $(\mu_3, \mu_4)$ are
displayed in Fig.~\ref{fig:KurtvsSkew} as red asterixes. For the sampling interval a value of
$\Delta x = 0.1 \, \mathrm{\mu m}$ has been chosen.

\vspace{3mm}

\begin{tabular}{c|c}
\hline\hline
\multicolumn{2}{l}{\textbf{Monte Carlo with varied}} \\
\multicolumn{2}{l}{\textbf{skewness and kurtosis}} \\
\hline\hline
$s(R_\mathrm{k})$ / $\mathrm{n m}$ & $48 \pm 3$ \\
$s(R_\mathrm{pk})$ / $\mathrm{n m}$ & $27 \pm 2$ \\
$s(R_\mathrm{vk})$ / $\mathrm{n m}$ & $50 \pm 4$ \\
\hline
$s$(Mr1) / \%  & $0.56 \pm 0.04$ \\
$s$(Mr2) / \%  & $0.80 \pm 0.06$\\
\hline\hline
\end{tabular}

\vspace{3mm}

With a variation of the statistical moments we could take influence on the resulting standard deviation pushing it up
to the values of the experimental result revealing that for any engineering process a texture assessment on a prototype
is required.
For fixed third and fourth moment, just varying the second, the uncertainty of $R_\mathrm{k}$ shows the linear relation
to that of $R_\mathrm{q}$, since they are directly related as the \textsl{Abbott curve} is the cumulative \textsl{PDF}.
$s(R_\mathrm{pk})$ and $s(R_\mathrm{vk})$ are strongly influenced by the higher moments.

In contrast to experiment, the value of the $s(R_\mathrm{pk})$ remains smaller while
$s(R_\mathrm{vk})$ is reproduced well, revealing that the Johnson functions
represent the left tail well enough but not the right side of the \textsl{PDF}. This shows that the proposed method
gives an approximate estimate, but that a detailed analysis and a precise uncertainty estimate requires a
more complex model of the stochastic processes.

\section{Conclusion}

Deriving the standard deviation of material ratio parameters caused by the inhomogeneity of
surface textures can be approximated coarsely from a single scan.
A Monte Carlo method that employs the autocorrelation length of the scanned profile and the first four
statistical moments of its amplitude distribution has been proposed. It is based on a model autocorrelation
function, an exponential or Gaussian, parameterized by the experimental autocorrelation length and on a model
probability density function of the Johnson system of which the parameterization is derived from the experimental
statistical moments.

Employing only one single profile has the advantage that the procedure can well be implemented into roughness
analysis software without any additional statistics information. Our investigations comparing
Monte Carlo with a high statistics experiment have shown that this may underestimate the value of the standard
deviation. To assess the uncertainty more precisely, more statistics is required, which can well be obtained by
scanning more profiles that are irregularly distributed across the surface delivering a
greater variation of the Monte Carlo generated profiles.
% The set of several profiles
%delivers a distribution of statistical moments to generate different probability density functions delivering a
%greater variation of the Monte Carlo generated profiles. 
A future goal is to develop more complex but still feasible model of surface texture statistics and
a learning system filling a data base for different classes of topographies.\\
~\\
% May be the parameters characterizing surfaces together with
%their manufacturing process parameters could be included to the \textsl{Standard Reference Data SRD} of NIST
%or a similar data base.
% http://www.nist.gov/srd/nist-databases-indexed-by-discipline.cfm

We are grateful to the anonymous referees to having taken their time for thoroughly reviewing this article.\\
~\\

\bibliography{../../utils/surftec,../../utils/qspm,../../utils/math}
\bibliographystyle{unsrturl}

\end{document}